\documentclass[prl,twocolumn,floatfix,superscriptaddress,longbibliography,nofootinbib]{revtex4-1}
\RequirePackage[sort&compress]{natbib}

\usepackage{amssymb,amsfonts,amsmath}
\usepackage{bm}
\usepackage[pdftex]{graphicx}
\usepackage{xcolor}
\usepackage{setspace}
\usepackage{subfigure}
\usepackage{comment}

\usepackage{hyperref}

\usepackage{soul}

\definecolor{darkGray}{RGB}{153,153,153}
\definecolor{darkBlue}{RGB}{37,113,161}
\definecolor{darkGreen}{RGB}{113,161,37}
\definecolor{darkRed}{RGB}{186,21,24}

\newcommand{\ket}[1]{\mbox{$|#1\rangle$}}

\usepackage{booktabs}

\newcommand{\heading}[1]{ \vspace{0.25truecm}\hspace{0.25truecm}\emph{#1}.--~}
\usepackage{textcomp}

\DeclareMathOperator{\Tr}{Tr}

\begin{document}

\title{Maximum entropy network states for coalescence processes}

\author{Arsham Ghavasieh}
\affiliation{Department of Physics, University of Trento, Via Sommarive 14, 38123 Povo (TN), Italy}

\author{Manlio de Domenico}
\email[]{manlio.dedomenico@unipd.it}
\affiliation{Department of Physics and Astronomy "Galileo Galilei", University of Padua, Via F. Marzolo 8, 315126 Padova, Italy}
 
\date{\today}

\begin{abstract}
Complex network states are characterized by the interplay between system's structure and dynamics. One way to represent such states is by means of network density matrices, whose von Neumann entropy characterizes the number of distinct  microstates compatible with given topology and dynamical evolution. 
In this Letter, we propose a maximum entropy principle to characterize network states for systems with heterogeneous, generally correlated, connectivity patterns and non-trivial dynamics. We focus on three distinct coalescence processes, widely encountered in the analysis of empirical interconnected systems, and characterize their entropy and transitions between distinct dynamical regimes across distinct temporal scales. Our framework allows one to study the statistical physics of systems that aggregate, such as in transportation infrastructures serving the same geographic area, or correlate, such as inter-brain synchrony arising in organisms that socially interact, and active matter that swarm or synchronize.
\end{abstract}

\date{\today}

\maketitle


The pioneering work by Jaynes~\cite{jaynes1957information,jaynes1957informationII} demonstrated how classical and quantum mechanics are intimely related to information theory and its quantum generalization~\cite{vedral2002role}. For complex systems with an interconnected structure, usually represented by a complex network of interactions between system's units~\cite{boccaletti2006complex}, often coupled to~\cite{dedomenico2013mathematical} or interdependent with other systems~\cite{buldyrev2010catastrophic}, such a relationship is still subject of investigation. It has been proposed to encode network states into operators resembling density matrices usually adopted in quantum statistical mechanics to represent entangled systems~\cite{braunstein2006laplacian,anand2011shannon,de2015structural} and to use the definition of von Neumann entropy to characterize their information content with applications ranging from cellular biology to connectomics and machine learning~\cite{ghavasieh_SARSCOV2,nicolini2020scale,baccini2022weighted}. However, only recently such operators have been defined in such a way that some fundamental properties, such as sub-additivity and extensivity, are satisfied~\cite{de2016spectral}. Furthermore, while network response to structural and dynamical perturbations — e.g., from random failures and targeted attacks leading to system’s disintegration into sub-systems to disruption of information flow –has been widely studied~\cite{albert2000error,callaway2000network,dorogovtsev2008critical,bashan2013extreme,morone2015influence,osat2017optimal,grassia2021machine}, 
the reverse process in which multiple sub-systems coalesce into a larger structural or functional whole and exhibit emergent properties  --- e.g., multicellular organisms formed from combinations of cells or neural systems from distinct individuals that collectively respond to shared external stimuli --- remains largely unknown.

In this Letter, we explore how information dynamics in systems emerging from the coalescence of sub-systems is constrained by new laws of additivity and extensivity. To this aim, we investigate three types of coalescence processes frequently observed in empirical interconnected systems, from biological networks to transportation infrastructures, in terms of density matrices obtained from a maximum entropy principle. We consider (i) aggregation, where two sub-systems with the same units integrate their connectivity patterns--- e.g., joining two air carriers serving the same region; ii) combination, where two sub-systems with different units form a network from the union of units and connections--- e.g., task-specific segregated circuits becoming part of a more complex network; iii) correlation, where two separate sub-systems interact with the same environment and their states become correlated--- e.g., a fish school facing a predator where separated neural systems of fishes collectively responds to the shared environment, as well as swarm and synchronizing organisms~\cite{mirollo1990synchronization,couzin2005effective,gomez2022intermittent}. We study how -- by adding units, connections or correlations -- sub-systems and interdependence affects the network state of the whole and its corresponding von Neumann entropy, to quantify the effect of coalescence on diversity of communication pathways between units. For some cases of interest, we demonstrate under which conditions entropy is sub-additive and show that different behaviors emerge, including phase transitions from sub-extensive to super-extensive regimes.

\heading{Maximum-entropy principle for network states} Let us consider a system with structure represented by a graph $\mathcal{G}(V, E)$, with $N=|V|$ indicating the number of system's units and $2|E|$ indicating the number of connections. The connectivity pattern is mathematically represented by the adjacency matrix $\mathbf{W}$, where the entry $W_{ij}$ ($i,j=1,2,...,N$) is 0 if there is no connection between nodes $i$ and $j$ and it is a positive number otherwise.

Furthermore, let us assume that it is possible to encode information about a network state into an operator $\boldsymbol{\rho}$, that we name density matrix in the following.

The operator $\boldsymbol{\rho}$ should satisfy the following properties: (A1) it is a continuous function of the adjacency matrix:  $\boldsymbol{\rho}=\varphi(\textbf{W})$; (A2) it is positive semi-definite; (A3) it generalizes the concept of probability to networks, i.e., $\Tr{\boldsymbol{\rho}}=1$, similarly to the density matrix in quantum mechanics~\cite{fano1957description,mcweeny1960some}.

Let us introduce a function $S(\boldsymbol{\rho})$ to measure the entropy of the network. As for classical and quantum entropy~\cite{jaynes1957information,jaynes1957informationII,jizba2019maximum}, we require some properties for $S$ to be well defined, with the additional requirement that its value cannot depend on how one labels the network units. Formally, we ask for: (B1) Positivity: $S(\boldsymbol{\rho})\geq 0$; 
(B2) Invariance under node relabeling: $S(\boldsymbol{\rho}')=S(\boldsymbol{\rho})$ if $\boldsymbol{\rho}'=f(\Pi~W~\Pi^{-1})$, being $\Pi$ a permutation matrix or a combination of permutation matrices;
(B3) Continuity: $S(\cdot)$ is a continuous function of $\boldsymbol{\rho}$;
(B4) Sub-additivity: $S(\boldsymbol{\rho}_{A+B}) \leq S(\boldsymbol{\rho}_{A}) + S(\boldsymbol{\rho}_{B})$.

Following von Neumann and Jaynes, it is possible to show that the function
\begin{eqnarray}
\label{def:entropy}
S(\boldsymbol{\rho}) = -\Tr{[\boldsymbol{\rho}\log_2 \boldsymbol{\rho}]}
\end{eqnarray}
satisfies (B1)--(B4). 
Note that if $\{\eta_{k}\}$ are the eigenvalues of $\boldsymbol{\rho}$, then $S(\boldsymbol{\rho})=-\sum_{k}\eta_{k}\log_2 \eta_{k}$, which is equivalent to the Shannon entropy of the eigenvalues distribution. In the same spirit of Jaynes' pioneering work on information theory and statistical mechanics~\cite{jaynes1957informationII}, it is possible to define a maximum entropy principle (MEP) in the context of network states. 

Let $f(\textbf{W})$ be a generic function  of the adjacency matrix, representing a network $\mathcal{G}(V,E)$. Since the Kullback-Leibler divergence between two density matrices $\boldsymbol{\rho}$ and $\boldsymbol{\sigma}$ is non-negative, 
it follows that 
\begin{eqnarray}
\label{diseq:lemma}
\Tr{[\boldsymbol{\rho}\log_{2}\boldsymbol{\rho}]} \geq \Tr{[ \boldsymbol{\rho}\log_{2}\boldsymbol{\sigma})]}.
\end{eqnarray}
By definition $\Tr{\boldsymbol{\rho}}=1$, and let us add an additional constrain to the expected value of  $f(\textbf{W})$ by requiring that $\Tr{[\boldsymbol{\rho}f(\textbf{W})]}=C$, being $C$ a constant scalar. Provided that $\boldsymbol{\sigma}$ can be any network density matrix, let us consider the specific one defined by
\begin{eqnarray}
\label{def:maxent-matrix}
\boldsymbol{\sigma}=\frac{e^{-\tau f(\textbf{W})}}{\Tr{[e^{-\tau f(\textbf{W})}}]}
\end{eqnarray}
that, after substitution into Eq.\,(\ref{diseq:lemma}) and some algebra leads to
\begin{eqnarray}
\label{eq:ineq}
S(\boldsymbol{\rho}) \leq \frac{\tau}{\log2} C + \log_{2} Z,
\end{eqnarray}
where the right-hand side terms do not depend on $\boldsymbol{\rho}$ and $Z=\Tr{[e^{-\tau f(\textbf{W})}}]$. Since one can vary $\boldsymbol{\rho}$ in the space of all network density matrices satisfying the same requirements, the corresponding entropy $S(\boldsymbol{\rho})$ will be upper bounded as in Eq.~(\ref{eq:ineq}) regardless of such a choice, with equality holding only if $\boldsymbol{\rho}=\boldsymbol{\sigma}$. We can therefore conclude that Eq.\,(\ref{def:maxent-matrix}) defines the network density matrix which maximizes the entropy function defined by Eq.\,(\ref{def:entropy}). 

Provided that $\tau$ indicates a time parameter, the above result has an elegant interpretation in the context of complex networks: the density matrix $\boldsymbol{\rho}$ representing a maximum-entropy network state (MENS) corresponds to the propagator of a linear dynamics on the top of the system. Such a dynamics usually appears close to the meta-stable state or at equilibrium. By indicating with $\ket{\psi(\tau)}$ such a state, the system's linear response to a small perturbation to $\ket{\psi(\tau)}$ is governed by the equation
\begin{eqnarray}\label{eq:lin_dyn}
\frac{\partial \ket{\psi(\tau)}}{\partial \tau} \approx -\mathbf{H}\ket{\psi(\tau)}.
\end{eqnarray}
Here the control operator is given by $\mathbf{H}\approx f(\textbf{W})$ and $\tau$ indicates the time scale at which the perturbation propagates among nodes. Note that $\boldsymbol{\sigma}$ and $Z$ resemble, respectively, the thermal state and partition function in quantum statistical physics, when $\mathbf{H}$ is a Hermitian matrix. In fact, for linear dynamics and a uniform distribution of perturbations over nodes, the same density matrix can be derived from Eq.~(\ref{eq:lin_dyn}) describing signal propagation in complex networks.

It is worth remarking that when $\mathbf{H}$ is the combinatorial Laplacian of an undirected graph, the above properties hold. In other cases, such as for random walks on networks~\cite{noh2004random,burda2009localization} this is not the case in general, since the normalized Laplacian operator might not be Hermitian. However, in the case of classical random walks~\cite{noh2004random}, it is possible to perform a transformation of the eigenvectors that leads to a symmetric normalized Laplacian with the same eigenvalues of the non-Hermitian one, corresponding to the generator of a valid quantum walk~\cite{faccin2013degree,biamonte2019complex} and leading to a MENS.

\heading{Variational method} The same result should be obtained from a variational method by adequately constraining the functional
\begin{eqnarray}
\mathcal{L}&\equiv& S(\boldsymbol{\rho})-\lambda_{0}(\Tr{\boldsymbol{\rho}}-1) - \sum_{k=1}^{n}\lambda_{k}(\Tr{[\boldsymbol{\rho}f_{k}(\textbf{W})]}-C_{k}),\nonumber
\end{eqnarray}
where $C_{k}=\Tr{[\boldsymbol{\rho} f_{k}(\mathbf{W})]}$, with $k=1,...,n$ are constraints to be satisfied simultaneously, while $\lambda_{0}$ and $\lambda_{k}$ are Lagrangian multipliers. In this case we can think about the control operator as $\mathbf{H}\approx \sum_{k}f_{k}(\textbf{W})$. The network state which satisfies the constraints is
given by
\begin{eqnarray}
\boldsymbol{\rho} =  \frac{e^{-\sum\limits_{k=1}^{n}\lambda_{k}^{\star} f_{k}(\mathbf{W})}}{Z},
\end{eqnarray}
where $\lambda_{k}^{\star}=\lambda_{k}\log 2$ and
$Z=\Tr{\exp{\left[-\sum\limits_{k=1}^{n}\lambda_{k}^{\star} f_{k}(\mathbf{W})\right]}}$,
and $\lambda_{k}$ is determined by solving $-\frac{\partial \log Z}{\partial \lambda_{k}} = C_{k}$.

It is worth wondering why choosing the density matrix that maximizes the entropy is correct, since any other matrix satisfying the same properties would be suitable as well. In fact, if we interpret the entropy $S(\boldsymbol{\rho})$ in Shannon's terms, i.e., as a measure of information, then it is desirable to use the network state that simultaneously satisfies the properties (A1)--(A3) and the constraints, in absence of further information about the system.


First, let us focus on the simplest case with only one constraint ($n=1$), so that $Z=\Tr{e^{-\lambda f(\mathbf{W})}}$, where $\lambda=\lambda_{1}^{\star}$ and $f=f_1$ for simplicity. Let us write $f(\mathbf{W})=\tau' f'(\mathbf{W})$, being $\tau'$ a real-valued parameter. From the constraint $\Tr{[\boldsymbol{\rho} f(\mathbf{W})]} = C$ it follows that 
\begin{eqnarray}
\Tr{\left[\frac{e^{-\lambda \tau' f'(\mathbf{W})}}{Z(\lambda,\tau')} \tau' f'(\mathbf{W})\right]} = \tau' C.\nonumber
\end{eqnarray}
Note that by introducing the parameter $\tau=\lambda \tau'$ the Lagrange multiplier is absorbed and we obtain the maximum entropy density matrix
\begin{eqnarray}
\boldsymbol{\rho}(\tau) = \frac{e^{-\tau f'(\mathbf{W})}}{Z(\tau)},\qquad Z(\tau)=\Tr{e^{-\tau f'(\mathbf{W})}},\nonumber
\end{eqnarray}
under the constraint $\Tr{\left[\boldsymbol{\rho} f'(\mathbf{W})\right]} = C$, a result which agrees, exactly, with the one obtained by starting from the Kullback-Leibler divergence. Remarkably, this network state has been used in the literature~\cite{de2016spectral,ghavasieh2020statistical} to quantify the relative importance of units for the structural and functional robustness of empirical systems from the human connectome to the power grids~\cite{ghavasieh_Structural_robustness,ghavasieh_functional_robustness}, identify mesoscale functional organization in fungal networks~\cite{ghavasieh_fungal} and communication in the healthy and pathological human brains~\cite{nicolini2020scale,benigni2021persistence}, characterize  protein networks in virus-host interactions~\cite{ghavasieh_SARSCOV2}, reduce the dimensionality of multiplex systems~\cite{ghavasieh2020enhancing}, identify cores responsible for controlling information flow~\cite{Villegas2022laplacian} and explore scale invariance through a renormalization group approach~\cite{villegas2022rg}.

\begin{figure}[!th]
\centering
\includegraphics[width=.5\textwidth]{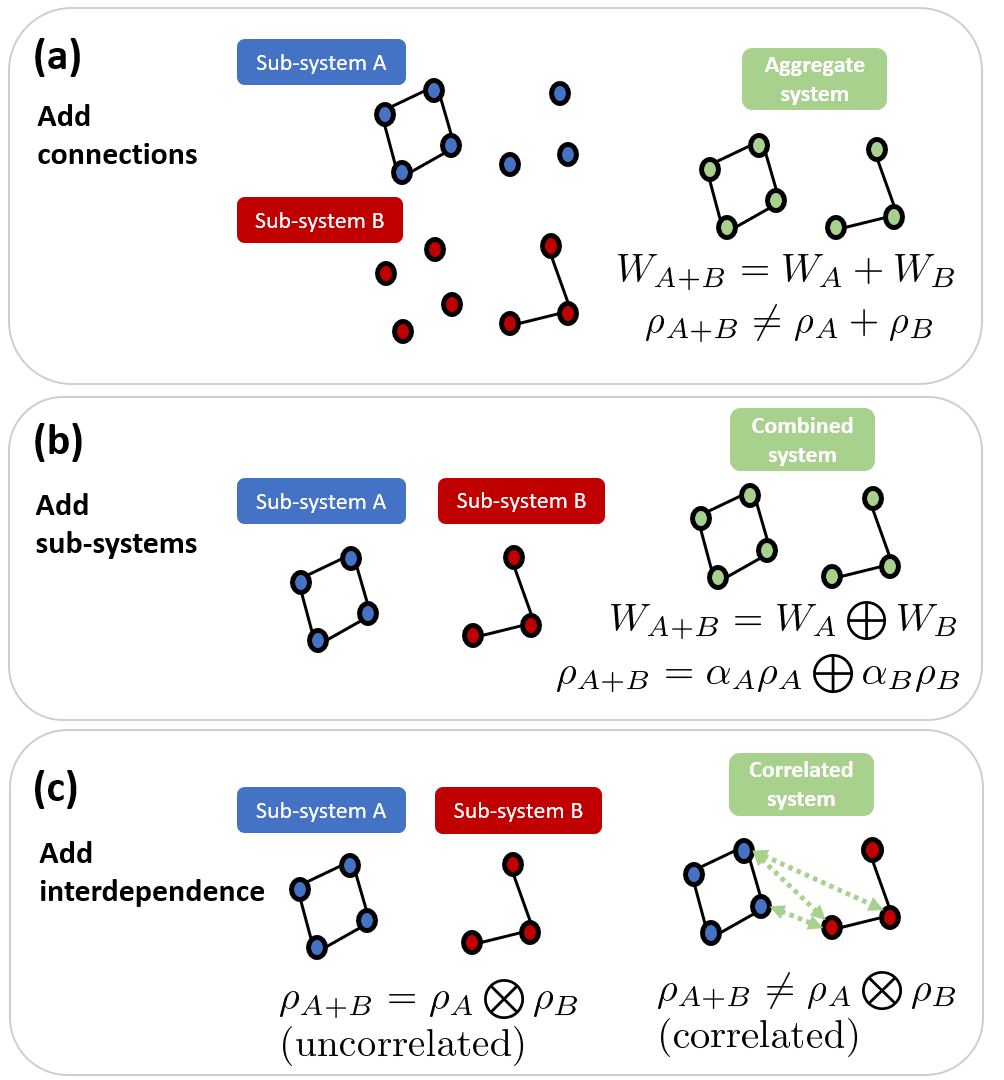}
\caption{\label{fig:schematic}\textbf{Schematic of sub-systems coalescencing into a complex network.} (a) Aggregation: two sub-systems defined over the same set of units (nodes) with distinct connectivity patterns (link) which are merged into a single network. (b) Combination: two sub-systems defined over different units are combined into a network consisting of the union of their nodes and links. (c) Correlation: two distinct sub-systems interact with a shared environment, adding new interdependent relationships among units. Here, $\mathbf{W}$ and $\boldsymbol{\rho}$ represent, respectively, the adjacency matrix and the density matrix (network state).  See the text for further details.}
\end{figure}

Let us focus on another case, where $f'(\mathbf{W})=\mathbf{H}+\mathbf{H}^{\dag}$, being $\mathbf{H}$ the control operator obtained by some linear function of $\mathbf{W}$. Under the additional requirement that $[\mathbf{H},\mathbf{H}^{\dag}]=0$ then
\begin{eqnarray}
\boldsymbol{\rho}(\tau) = \frac{e^{-\tau \mathbf{H}}e^{-\tau \mathbf{H}^{\dag}}}{Z(\tau)}, \qquad Z(\tau)=\Tr{[e^{-\tau \mathbf{H}}e^{-\tau \mathbf{H}^{\dag}}]}.\nonumber
\end{eqnarray}
Note that this result recovers a special case of a generalized density matrix~\cite{de_generalized_density_matrix_2022} where the dynamical process is linearized and perturbation probability is assumed to be uniform across nodes. 


\heading{Entropy of network states in coalescence processes} Let us now consider coalescence processes, where two initially disjoint sub-systems $A$ and $B$, represented by networks $\mathcal{G}_{A}(V_{A},E_{A})$ and $\mathcal{G}_{B}(V_{B},E_{B})$, respectively. Those two sub-systems can coalesce in different ways: here we consider (i) aggregation, (ii) combination and (iii) correlation. In the following we will consider the case of two sub-systems without loss of generality, remarking that the results can be easily extended to the case of a larger set os sub-systems.

\heading{(i) Aggregation} In aggregation processes, $V_A = V_B$ and the resulting network is obtained by adding up the connections of sub-systems $A$ and $B$ according to some rule $f(\mathbf{W}_{A}+\mathbf{W}_{B})$. An example of this process is related to the fusion of two air carriers serving the same region: there is no creation of new airports (nodes) and the routes served by the two companies are integrated. Let us use the following notation: $\mathbf{f}_{A+B} = f(\mathbf{W}_{A}+\mathbf{W}_{B})$, $\mathbf{f}_{A,B} = f(\mathbf{W}_{A,B})$. 
The coalesced system will be defined over the same state space as the two original sub-systems (Fig.~\ref{fig:schematic}A). It is possible to show that, in this case, the von Neumann entropy of the composite system is bounded by the sum of the entropy of the two sub-systems, i.e.: $S(\boldsymbol{\rho}_{A+B}) \leq S(\boldsymbol{\rho}_{A})+ S(\boldsymbol{\rho}_{B})$~\cite{de2016spectral, de_generalized_density_matrix_2022}, a behavior that can be defined as link sub-extensivity. After some algebra, we find that a general condition for link sub-extensivity to hold is given by the condition
\begin{eqnarray}
\langle\mathbf{f}_{A+B}\rangle - \langle\mathbf{f}_{A} + \mathbf{f}_{B}\rangle \leq \frac{1}{\tau}\log{\frac{Z_{A}Z_{B}}{Z_{A+B}}},\nonumber
\end{eqnarray}
where we have used the fact that $\langle\mathbf{f}_{X}\rangle=\Tr{\left[ \boldsymbol{\rho}_{X}\mathbf{f}_{X}\right]}$. If $\mathbf{f}_{A+B}=\mathbf{f}_{A}+\mathbf{f}_{B}$, the left-hand side of the above equation is identically zero and the condition to be satisfied reduces to $Z_{A+B}(\tau)\leq Z_{A}(\tau)Z_{B}(\tau)$. It can be shown that this condition is satisfied for any value of $\tau$: a special case is diffusion dynamics, governed by the combinatorial Laplacian, with a linear $f(\cdot)$ and sub-extensive entropy.

In the more general case that the aggregation mechanism is nonlinear, we define $\delta(\tau)=e^{\tau(\langle\mathbf{f}_{A+B}\rangle - \langle\mathbf{f}_{A} + \mathbf{f}_{B}\rangle)}$ and obtain the condition for link sub-extensivity is given by $X(\tau)\leq 1$, with
\begin{eqnarray}
X(\tau) &&= \delta(\tau)\frac{ Z_{A+B}(\tau)}{ Z_{A}(\tau)Z_{B}(\tau)}.
\end{eqnarray}
The behavior of $X(\tau)$ determines if there are phase transitions regarding extensivity. It clearly shows that link sub-extensivity depends on how system's and sub-systems' partitions functions compare. Also, sub-extensivity can be violated, even for small deviations from linearity, i.e. $\langle\mathbf{f}_{A+B}\rangle - \langle\mathbf{f}_{A} + \mathbf{f}_{B}\rangle \approx \epsilon(\tau)$, where $\delta(\tau)\approx e^{\epsilon(\tau)\tau}$ rapidly diverges ---e.g., for constant and positive $\epsilon(\tau)$--- leading to super-extensivity at some temporal scale $\tau_c$. Remarkably, when $\frac{ Z_{A+B}(\tau)}{ Z_{A}(\tau)Z_{B}(\tau)}<1$, nonlinear aggregation mechanisms are responsible for the existence of two distinct regimes: i) where the various layers are structurally decoupled and they act as independent entities ii) network layers are indistinguishable and the whole system behaves as a single-level network. Finally, instead of structural aggregation, it is possible to merge networks of the same size functionally. For instance, in case of multiplex networks with layers $A$ and $B$, random walk dynamics follows $f_{A+B} = (f_{A}+f_{B})/2$~\cite{ghavasieh2020enhancing}, leading to a negative $\epsilon(\tau)$ ensuring sub-extensivity at $\tau>\tau_c$. 

\heading{(ii) Combination} In combination processes, $V_A \neq V_B$ and the resulting network is obtained from the union of units and connections of sub-systems $A$ and $B$ (Fig.~\ref{fig:schematic}B). In this case, the adjacency matrix of the resulting network $\mathcal{G}_{A+B}$ is given by the direct sum $\mathbf{W}_{A}\bigoplus \mathbf{W}_{B}$, a a block diagonal matrix. It follows that $\mathbf{f}_{A\oplus B}=f(\mathbf{W}_{A}) \bigoplus f(\mathbf{W}_{B})$, where $f(\cdot)$ is some well-behaved function. For MENS the density matrix of the composite system is given by
\begin{eqnarray}
\boldsymbol{\rho}_{A+B}(\tau) &=&  \alpha_{A}\boldsymbol{\rho}_{A}(\tau) \bigoplus \alpha_{B} \boldsymbol{\rho}_{B}(\tau),
\end{eqnarray}
where $\alpha_{A}=Z_{A}/Z_{A+B}$, $\alpha_{B}=Z_{B}/Z_{A+B}$ and $Z_{A+B}= Z_{A} + Z_{B}$. Since $\alpha_{A} + \alpha_{B} = 1$, and the partition function can be related to the total energy of signalling in the network~\cite{de_generalized_density_matrix_2022}, $\alpha_{X}$ can be probabilistically interpreted as the fraction of signal energy in sub-system X. Therefore, we can estimate the information entropy of such fractions by $S_\alpha= - \alpha_{A}\log_{2}{\alpha_{A}} - \alpha_{B}\log_{2}{\alpha_{B}}$. It is straightforward to show that
\begin{eqnarray}
S_{A+B}=\alpha_{A} S_{A} + \alpha_{B} S_{B} + S_{\alpha},\nonumber
\end{eqnarray}
where $S_{\alpha}$ accounts for the uncertainty related to the location of signals in the combined system. For two coalescing sub-systems this quantity is at most 1~bit ($\alpha_A=\alpha_B=1/2$), whereas for $M$ sub-systems $\{A_{\ell}\}$ ($\ell=1,2,...,M$) it is at most $\log_{2}M$~bits. In this latter case, note that
\begin{eqnarray}
S_{\alpha} = S\left(\bigoplus_{\ell=1}^{M}\alpha_{\ell}\boldsymbol{\rho}_{\ell}(\tau)\right) - \sum_{\ell=1}^{M}\alpha_{\ell} S\left(\boldsymbol{\rho}_{\ell}(\tau)\right)
\end{eqnarray}
This result is interesting since it shows that by modeling the combination of  disjoint system as a single network we lose bits of information entropy due to increased uncertainty about its description. Note that we would lose this information if the involved sub-systems were isolated from each other. Note that this result is formally equivalent to the Holevo bound of a quantum communication channel~\cite{vedral2002role}, suggesting further investigation. 

If after combination we start to introduce new connections between the units belonging to distinct sub-systems, we will have the equivalent of an aggregation process, and von Neumann entropy is expected to exhibit link sub-extensivity or undergo a phase transition depending on the nonlinearity of the aggregating dynamics.


\heading{(iii) Correlation} In correlated processes, $V_A \neq V_B$ and the resulting network is obtained from considering the interdependencies between two sub-systems (Fig.~\ref{fig:schematic}C) that can emerge when they interact with a common environment.

For two independent quantum systems the density matrix of the composite system is separable, $\boldsymbol{\rho}_{A+B}=\boldsymbol{\rho}_{A} \bigotimes \boldsymbol{\rho}_{B}$, which is equivalent to physical systems without quantum correlations. However, for complex networks the meaning of such correlations is totally different, since the density matrix represents a superposition of information propagation patterns in the system~\cite{de_generalized_density_matrix_2022}. For instance, let us assume that a network is initially in a state $|\psi\rangle$--- often considered to be the steady state where the value of each node $j$ given by $\langle j|\psi\rangle$ does not change with time. If the $i$-th node is perturbed, the system evolves from $|\psi\rangle $ to another state $|\psi^{(i)}_{\tau}\rangle$ after $\tau$ time steps, and the perturbation propagation vector is given by $|\Delta \psi^{(i)}_{\tau}\rangle = |\psi^{(i)}_{\tau}\rangle - |\psi \rangle$. The same propagation can be encoded in a local propagator given by the outer product of propagation vector and its complex conjugate $\hat{\mathbf{U}}^{(i)} = |\Delta \psi^{(i)}_{\tau}\rangle \langle \Delta \psi^{(i)}_{\tau}|$
where the $j$-th diagonal element $\langle j|\hat{\mathbf{U}}^{(i)}|j\rangle$ gives the signal energy on top of node $j$, received from node $i$ at $\tau$. Also, the $jk$ off-diagonal element $\langle j|\hat{\mathbf{U}}^{(i)}|k\rangle$ encodes the correlation between node $j$ and $k$ in receiving signal amplitudes from $i$. Each perturbation appears with a probability ($p_i$ for node $i$) in a complex network, requiring a mathematical object storing a statistical description of propagation patterns: the statistical propagator $\hat{\mathbf{U}} =\sum\limits_{i}p_{i}\hat{\mathbf{U}}^{(i)}$. Finally, the generalized density matrix~\cite{de_generalized_density_matrix_2022} gives a statistical description of such a propagation, per unit of signal energy, and is given by $\boldsymbol{\rho}=\hat{\mathbf{U}}/Z$, where $Z=\Tr{\hat{\mathbf{U}}}$. Note that the MENS considered in previous sections is a special case of such a generalized density matrix. Here, the network state satisfies the same constraints of physical density matrices, being unit trace and positive semi-definite, and sub-additivity of the von Neumann applies. 
As mentioned before, the equality $S_{A+B} = S_{A} + S_{B}$ holds if the information flow in the two networks is independent--- e.g., two neural systems interacting with distinct environments. Yet, the propagation patterns can correlate. It can happen, for instance, when two separate networks interact with the same environment, like two individuals performing a common task that induce cross-correlations in the corresponding neural activities~\cite{sanger2012intra,liu2017inter,ahn2018interbrain,zhang2019correlated,kingsbury2019correlated}. In this case, observing the perturbation propagation in one network is highly informative about the other network, and the von Neumann entropy of the two correlated sub-systems is smaller than the summation of the single entropies.

\heading{Conclusion} Coalescence processes are frequently observed in empirical systems evolving to form larger networks from smaller units. We map those processes into operations with density matrices derived from a maximum entropy principle, and derive their entropy additivity and extensivity. This way, we distinguish processes tending to decrease systems' functional diversity--- i.e., the diversity of dynamical response to internal or external perturbations--- from super-additive ones, increasing the functional diversity. Remarkably, for nonlinear aggregation mechanisms, we predict a phase transition from sub- to super-extensivity depending on a certain temporal scale of signal propagation. For combination, the von Neumann entropy of the network equals the weighted entropy of its parts, plus the partition entropy encoding the uncertainty about the location of signals in sub-systems. For correlated sub-systems, sub-additivity behaves exactly like the one in quantum systems. Our results quantify the effect of systems integration on information dynamics, opening research directions on the relationship between network and quantum information theory.

\bibliographystyle{apsrev4-1}
\bibliography{draft}

\end{document}